\documentclass[letterpaper,twocolumn,showpacs,prl,aps]{revtex4-1}

\usepackage{amsmath}  
\usepackage{amsfonts} 
\usepackage{graphicx} 
\usepackage{amssymb}

\begin{document}

\title{Space charge and screening in bilayer graphene}

\author{Eugene B. Kolomeisky$^{1}$, Joseph P. Straley$^{2}$, and Daniel L. Abrams$^{1}$}

\affiliation
{$^{1}$Department of Physics, University of Virginia, P. O. Box 400714,
Charlottesville, Virginia 22904-4714, USA\\
$^{2}$Department of Physics and Astronomy, University of Kentucky,
Lexington, Kentucky 40506-0055, USA}

\date{\today}

\begin{abstract}
Undoped bilayer graphene is a two-dimensional semimetal with a low-energy excitation spectrum that is parabolic in the momentum.  As a result, the screening of an arbitrary external charge $Ze$ is accompanied by a reconstruction of the ground state:  valence band electrons (for $Z>0$) are promoted to form a space charge around the charge while the holes leave the physical picture.  The outcome is flat neutral object resembling the regular atom except that for $Z \gg 1$ it is described by a strictly linear Thomas-Fermi theory. This theory also predicts that the bilayer's static dielectric constant is the same as  that of a two-dimensional electron gas in the long-wavelength limit.
\end{abstract}


\maketitle

The experimental isolation of graphene, a single monolayer of graphite, has opened a new chapter in the physics of interacting electrons in condensed matter \cite{graphene_review}.  Graphene is a two-dimensional semimetal whose low-energy elementary excitations obey a pseudo-relativistic dispersion law.  This relates graphene's physics to the world of quantum electrodynamics (QED).  The counterpart of the limiting velocity is two orders of magnitude smaller than the speed of light in vacuum, which amplifies the effects of interactions and makes a variety of strong-coupling QED-like phenomena experimentally accessible.  One such effect that has received considerable attention \cite{ZP},  "atomic collapse",  is an instability (in three dimensions) of the physical vacuum which is induced by the very large electric field of a nucleus of electric charge $Ze$ with $Z>Z_{c}\approx 170$.  The outcome of the instability is the creation of electron-positron pairs with the electrons screening the nuclear charge and the positrons leaving the physical picture.  A similar phenomenon of screening by space charge takes place in graphene, with the electron and hole excitations substituting for the electron-positron pairs, except that $Z_{c}\simeq1$ (for suspended graphene)  \cite{Shytov} makes the effect detectable \cite{collapse}.  Effects analogous to QED's atomic collapse with experimentally accessible $Z_{c}$ are also predicted to take place in narrow band-gap semiconductors and Weyl semimetals \cite{KSZ}.       

Bilayer graphene (two Bernal-stacked graphene monolayers) has recently attracted a great deal of attention \cite{bireview}.  Some of its electronic properties resemble those of the monolayer while other are even more unusual, because the bilayer is a two-dimensional semimetal with an unusual band structure.  The goal of this paper is to clarify the physics of the screening response of the bilayer graphene.  Despite the presence of the effect of atomic collapse for $Z\geqslant Z_{c}$, the ground state of the undoped bilayer is found to be unstable with respect to creation of a screening space charge already at $Z=1$.  The $1\ll Z \ll Z_{c}$ response is described by a Thomas-Fermi (TF) theory \cite{LL3} that while being "accidentally" linear also predicts that the static dielectric function of the material is the same as that of a two-dimensional electron gas in the long-wavelength limit \cite{Stern}.  This disagrees slightly with the result of the calculation based on the random-phase approximation (RPA) \cite{Hwang}.  We additionally find that the system responds to an external potential of given charge by creating around it a space charge of the same magnitude and opposite sign;  the response to any shorter-range potential always involves both electrons and holes of zero net charge.        

The low- to intermediate-energy elementary excitations in bilayer graphene obey a hyperbolic dispersion law \cite{dispersion,bireview} 
\begin{equation}
\label{dispersion_law}
\varepsilon(\textbf{p})=\pm mv^{2}\left (\sqrt{1+\left (\frac{p}{mv}\right )^{2}}-1\right )
\end{equation}
where $\textbf{p}$ is the two-dimensional momentum vector, the upper and lower signs correspond to the conduction and valence bands, respectively, $m$ is the effective electron mass, and $v$ is the Fermi velocity.  In the undoped case that we consider, the conduction band is empty while the valence band is occupied in the absence of a perturbing potential.  The conduction and valence bands in (\ref{dispersion_law}) meet at $\textbf{p}=0$ in a parabolic fashion $\varepsilon=\pm p^{2}/2m$, where $m\approx(0.03\div0.05) m_{0}$ ($m_{0}$ is the electron mass in vacuum) \cite{bireview,dispersion}.  For larger momenta the spectrum (\ref{dispersion_law}) becomes linear $\varepsilon=\pm vp$, coinciding with that of the monolayer graphene; $v \approx c/300$ \cite{graphene_review}.  The crossover between the parabolic and linear behavior occurs at $p\approx mv$ corresponding to a carrier density $n^{*}\approx 4 \times 10^{12} cm^{-2}$;  the two-band approximation (\ref{dispersion_law}) is applicable up to the carrier density of $8n^{*}$ which is when the next available band can become occupied \cite{bireview,dispersion}.  The dispersion law (\ref{dispersion_law}) also neglects small (and for us unimportant) trigonal warping effects due to the underlying symmetry of the graphene lattice \cite{bireview,dispersion}.  

When combined with the uncertainty principle, the existence of the limiting velocity $v$ in Eq.(\ref{dispersion_law}) dictates that the electron is delocalized on the scale given by the counterpart of the Compton wavelength $\lambda=\hbar/mv$.  Since the bilayer electrons are both "light" ($m\ll m_{0}$) and "slow" ($v\ll c$), $\lambda$ is of the order of a nanometer, thus significantly exceeding both the lattice constant and the separation between monolayers \cite{bireview,dispersion}.  Then in the low-energy limit we can use a continuum theory that treats the bilayer as a planar structureless system characterized by a background dielectric constant $\kappa$ that is due to the bilayer's own electrons and the  polarization of the substrate.  

We begin by looking at the binding properties of an external potential of net charge $Ze$, $\varphi_{ext}(r\rightarrow \infty) \rightarrow Ze/\kappa r$, (where $r$ is the in-plane position) which does not introduce an asymmetry between individual monolayers \cite{bireview, dispersion}.  Such a potential could be due to dopants or can be created in a controlled manner by two gates symmetrically positioned on both sides of the bilayer.  Following Ref.\cite{KSZ}, we estimate the ground state properties starting from   the classical expression for the energy of an electron in the field of positive point charge $Ze$ placed within the bilayer plane:  
\begin{equation}
\label{Cl_energy}
\varepsilon = mv^{2}\left (\sqrt{1+\left (\frac{p}{mv}\right )^{2}}-1\right )-\frac{Ze^{2}}{\kappa r}
\end{equation}
Since the electron position cannot be determined to better accuracy than $\hbar$ divided by the uncertainty of momentum, $p$ and $r\gtrsim \hbar/p$ entering Eq.(\ref{Cl_energy}) may be regarded as the typical momentum and size of the quantum state, respectively.  Then the state energy  can be estimated as
\begin{equation}
\label{energy_vs_p}
\varepsilon(p)\gtrsim  mv^{2}\left (\sqrt{1+\left (\frac{p}{mv}\right )^{2}}-1-\frac{Z\alpha p}{mv}\right ), \alpha=\frac{e^{2}}{\kappa \hbar v}
\end{equation}
where $\alpha$ is the material fine structure constant.  Minimizing with respect to the free parameter $p/mv$ we find 
\begin{eqnarray}
\label{gs_properties}
p_{0}&\simeq& mv\frac{Z\alpha}{\sqrt{1-(Z\alpha)^{2}}},~~~~r_{0}\simeq \lambda\frac{\sqrt{1-(Z\alpha)^{2}}}{Z\alpha},\nonumber\\
\varepsilon_{0}& \simeq & mv^{2}\left (\sqrt{1-(Z\alpha)^{2}}-1\right )
\end{eqnarray}  
In the "non-relativistic" limit $Z\alpha \ll 1$ the size of the ground state and the ground-state energy become
\begin{equation}
\label{non-relativistic}
r_{0} \simeq \frac{b}{Z},\varepsilon_{0}\simeq -Z^{2}Ry^{*}, b=\frac{\kappa \hbar^{2}}{me^{2}}\equiv \frac{\lambda}{\alpha}, Ry^{*}=\frac{\hbar^{2}}{2mb^{2}}
\end{equation}
These expressions are close counterparts to the ground-state properties of a hydrogen-like ion of nuclear charge $Ze$, except that here the electron is confined to the plane;  the length scale $b$ and the energy scale $Ry^{*}$ are the bilayer Bohr radius and the effective Rydberg, respectively.

These arguments also predict that the minimum of (\ref{energy_vs_p}) only exists for $Z<Z_{c}\simeq 1/\alpha$, the onset of atomic collapse \cite{ZP}:  as $Z\rightarrow Z_{c}-0$, the ground state becomes sharply localized ($r_{0}\rightarrow 0$), the typical electron momentum diverges ($p_{0}\rightarrow \infty$), and the ground-state energy approaches $-mv^{2}$.  Since the atomic collapse is an "ultra-relativistic" effect, the exact value of the critical charge must be the same as that for  monolayer graphene $Z_{c}=1/(2\alpha)$ \cite{Shytov} (thus substantiating our estimate $Z_{c}\alpha \simeq 1$), as the energy spectra of the systems coincide for large momenta.  As in the case of monolayer graphene one finds  $Z_{c}\simeq \kappa$ and $b\simeq \kappa \times nm$.  The physics in the $Z>Z_{c}$ regime is similar to that in the monolayer graphene \cite{Shytov,graphene_TF} and will not be discussed here.  Instead we focus on the $Z<Z_{c}$ case where the bilayer's response is qualitatively different.  

The key observation is that the bilayer ground state in the presence of an external potential is \textit{always} unstable with respect to creation of space charge.  Indeed, let us start with charge $Ze$ and no electrons present.  Since the band gap is zero and the single-particle ground-state energy $\varepsilon_{0}$, Eq.(\ref{gs_properties}), is negative, the system energy can be lowered by creating an electron-hole pair, binding the electron to the charge and removing the hole to infinity.  Unless $Z=1$, the buildup of the screening charge via the same mechanism continues until neutrality is reached.  The outcome is the 2D counterpart of a regular three-dimensional atom with $Z$ electrons bound to the external charge.   For  $Z\gg 1$, the number of electrons is large and  TF theory can be used to compute "atom" properties;  this is also a theory of bilayer's screening response.  The TF theory is known to recover the exact quantum-mechanical description in the $Z\rightarrow \infty$  limit \cite{Lieb} and to be reliable for slowly varying external potentials \cite{AM}.   

The central object of the TF theory is the self-consistent or total potential $\varphi(\textbf{r})$ felt by an electron at position $\textbf{r}$ which is due to the external potential $\varphi_{ext}(\textbf{r})$ and the remaining electrons of the screening cloud:
 \begin{equation}
\label{scpotential}
\varphi(\textbf{r})=\varphi_{ext}(\textbf{r})-\frac{e}{\kappa}\int \frac{n(\textbf{r}')d^{2}r'}{|\textbf{r}-\textbf{r}'|}
\end{equation}
where $n(\textbf{r})>0$ is the electron number density within the bilayer.  In the undoped case the electron chemical potential is zero which leads to the condition of equilibrium
\begin{equation}
\label{chem_potentials}
\frac{p_{F}^{2}}{2m}-e\varphi=0
\end{equation}
where $p_{F}(\textbf{r})$ is the local Fermi momentum.  In writing the kinetic energy term in (\ref{chem_potentials}) in the "non-relativistic" form we made an additional simplification (that can be justified for $\kappa \gg1$), approximating the hyperbolic dispersion law (\ref{dispersion_law}) by a parabolic one.  Eliminating the Fermi momentum from Eq.(\ref{chem_potentials}) via $p_{F}^{2}(\textbf{r})=4\pi \hbar^{2}n(\textbf{r})/g$ ($g=4$ is the degeneracy factor that accounts for the spin and valley degrees of freedom of the electrons in graphene) gives a linear relationship between the number density of the space charge and the total potential
\begin{equation}
\label{n_of_phi}
n(\textbf{r})= \frac{mge}{2\pi \hbar^{2}}\varphi(\textbf{r})
\end{equation} 
Substitution of Eq.(\ref{n_of_phi}) into Eq.(\ref{scpotential}) leads to the central equation of the TF theory:
\begin{equation}
\label{TF_equation}
\varphi(\textbf{r})=\varphi_{ext}(\textbf{r})-\frac{q_{s}}{2\pi}\int\frac{\varphi(\textbf{r}')d^{2}r'}{|\textbf{r}-\textbf{r}'|},~~~~~q_{s}=\frac{g}{b}
\end{equation}
The linearity of this equation is a special feature of the two-dimensionality of the electrons and the parabolic character of their low-energy dispersion law;  it is an integral equation because the two-dimensional electrons interact according to the three-dimensional Coulomb law.   

The discussion so far has assumed that $\varphi_{ext}>0$ causes a material response in the form of screening by electrons with number density $n>0$ that reside where $\varphi>0$.  However, the final equations (\ref{n_of_phi}) and (\ref{TF_equation}) also apply to holes ($n<0$) that could be created by $\varphi_{ext}<0$ and would reside in the regions of space where $\varphi<0$.

Taking the Fourier transform of Eq.(\ref{TF_equation}) reduces it to an algebraic equation with solution
\begin{equation}
\label{Fourier_solution}
\varphi(\textbf{k})=\frac{\varphi_{ext}(\textbf{k})}{1+q_{s}/k}
\end{equation}
where $\varphi(\textbf{k})$ and $\varphi_{ext}(\textbf{k})$ are the Fourier transforms of the potentials.  We now see that the parameter $q_{s}$ introduced in Eq.(\ref{TF_equation}) is the screening wave vector:  the long-wavelength modes of the external potential satisfying $k\ll q_{s}$ are nearly completely screened while the short-wavelength modes, $k\gg q_{s}$, are hardly affected.  The linear relationship between the Fourier transforms of the external potential and total potential (\ref{Fourier_solution}) means that static dielectric function of the bilayer is 
\begin{equation}
\label{dielectric-constant}
\epsilon_{TF}(k)=\kappa \left (1+\frac{q_{s}}{k}\right )
\end{equation}
(the subscript "TF" refers to the TF approximation).  This has the form of the dielectric function of a two-dimensional electron gas in the long-wavelength limit \cite{Stern} but differs from the RPA result \cite{Hwang,Nilsson}, for which the screening wavevector is larger by a factor of $\ln4 = 1.38..$.  The difference in the approaches is that RPA is a momentum-space perturbation theory which implicitly assumes that $e \varphi_{ext}(\textbf{k}) \ll \hbar^{2} k^{2} /m$, so that the single-particle band structure is only perturbatively affected \cite{perturbative}, whereas TF is a real-space approach most appropriate to the limit $e \varphi_{ext}(\textbf{k}) \gg \hbar^{2} k^{2}/m$, for which the valence band becomes partially filled (the ground state is simply different).  The TF limit is clearly more appropriate for the examples we will consider below.

Eqs.(\ref{n_of_phi})-(\ref{dielectric-constant}) can be alternatively derived starting from a standard two-dimensional TF energy functional
\begin{eqnarray}
\label{TF_functional}
E[n]&=&\frac{\pi \hbar^{2}}{mg}\int n^{2}(\textbf{r})d^{2}r+\frac{e^{2}}{2\kappa}\int\frac{n(\textbf{r})n(\textbf{r}')d^{2}rd^{2}r'}{|\textbf{r}-\textbf{r}'|}\nonumber\\
&-&e\int \varphi_{ext}(\textbf{r})n(\textbf{r})d^{2}r\nonumber\\
&=&\int \frac{d^{2}k}{(2\pi)^{2}}\left (\frac{\pi e^{2}}{\kappa^{2}q_{s}}\epsilon_{TF}(k)|n(\textbf{k})|^{2}-e\varphi_{ext}(-\textbf{k})n(\textbf{k})\right )\nonumber\\
\end{eqnarray}
Unconstrained minimization of the functional (\ref{TF_functional}) with respect to $n$ combined with the definition of the total potential (\ref{scpotential}) then reproduces Eqs.(\ref{n_of_phi})-(\ref{dielectric-constant}).  The Fourier representation of the functional (\ref{TF_functional}) also supplies an interpretation to the dielectric constant (\ref{dielectric-constant}) as being proportional to the stiffness coefficient characterizing the system's response to a charge density with wave vector $\textbf{k}$;  the inequality $\epsilon_{TF}(k)>1$ then implies (for $\varphi_{ext}=0$) stability of the semimetal ground state with respect to creation of  electron-hole pairs.  Substituting into Eq.(\ref{TF_functional}) its minimizer $n(\textbf{k})=\kappa^{2}q_{s}\varphi_{ext}(\textbf{k})/2\pi e\epsilon_{TF}(k)$ we arrive at the expression for the ground-state energy
\begin{equation}
\label{TF_gs_energy}
E_{0}=-\frac{\kappa^{2}q_{s}}{4\pi}\int\frac{d^{2}k}{(2\pi)^{2}}\frac{|\varphi_{ext}(\textbf{k})|^{2}}{\epsilon_{TF}(k)}
\end{equation}      

For an external potential with circular symmetry ($\varphi_{ext}(\textbf{r})=\varphi_{ext}(r)$), the Fourier inversion of Eq.(\ref{Fourier_solution}) produces a  real space solution in the form
\begin{equation}
\label{real_space solution}
\varphi(r)=\varphi_{ext}(\textbf{r})-\frac{q_{s}}{2\pi}\int_{0}^{\infty}\varphi_{ext}(k)J_{0}(kr)\frac{dk}{1+q_{s}/k}
\end{equation}
where $J_0(x)$ is the Bessel function.  By taking the $q_{s}\rightarrow \infty$ limit in Eq.(\ref{real_space solution}) and combining the outcome with Eq.(\ref{n_of_phi}) we find that $\varphi=0$ and
\begin{equation}
\label{classical}
n(r)=\frac{\kappa}{(2\pi)^{2}e}\int_{0}^{\infty}\varphi_{ext}(k)J_{0}(kr)k^{2}dk
\end{equation}
which is the classical electrostatics density distribution induced in a conducting plane by a circularly-symmetric external potential $\varphi_{ext}(k)$. 

When the external potential is that of a net charge $Ze$ (so that $\varphi_{ext}(r\rightarrow \infty) \rightarrow Ze/\kappa r$), one can take the $r\rightarrow \infty$ limit in Eq.(\ref{scpotential}) with the conclusions that the external potential is completely screened ($\int n(\textbf{r}) d^{2}r=Z$)  and that the total potential $\varphi$ falls off faster than $1/r$ at large $r$.  For a point charge $Ze$ within the bilayer plane one has $\varphi_{ext}(\textbf{k})=2\pi Ze/\kappa k$ and evaluation via Eq.(\ref{real_space solution}) gives a formula
\begin{equation}
\label{screened_Coulomb_real_space}
\varphi=\frac{Ze}{\kappa r} -\frac{\pi Zeq_{s}}{2\kappa}[\textbf{H}_{0}(q_{s}r)-Y_{0}(q_{s}r)]
\end{equation}        
for the screened Coulomb potential that is familiar from studies of the two-dimensional electron gas \cite{Stern} (here $\textbf{H}_0(x)$ and $Y_0(x)$ are the Struve function and Bessel function of the second kind, respectively).  For $q_{s}r\gg1$ we find $\varphi(r)\approx Zeq_{s}/\kappa(q_{s}r)^{3}$, a large distance decay typical of any external charge distribution of net charge $Ze$.  

Eqs.(\ref{screened_Coulomb_real_space}) and (\ref{n_of_phi}) give the potential and density distribution within a flat $Z\gg1$ "atom" with a point-like "nucleus".  Unlike an ordinary atom \cite{LL3}, the potential and density are strictly linearly proportional to $Z$; the characteristic length scale of the potential and density distributions is the Bohr radius $b$.  The TF result (\ref{screened_Coulomb_real_space}) is applicable as long as the electron de Broglie wavelength $2\pi \hbar/p_{F}(r)$ varies insignificantly with position $r$ which excludes both very small and very large distances:  $b/Z\ll r\ll bZ$.  This range is wider than that of the TF theory of the ordinary atom where the upper bound is  $r\simeq b$ \cite{LL3}.  It is straightforward to verify that practically all the electrons of the atom are confined within a radius $\sim bZ$, i.e. the size of such an atom grows linearly with $Z$, again in contrast to the ordinary atom where atomic sizes are of the order $b$ and approximately $Z$-independent.  Substituting $\varphi_{ext}(\textbf{k})=2\pi Ze/\kappa k$ into Eq.(\ref{TF_gs_energy}) one encounters a logarithmic divergence at large $k$,  which is an artifact of the continuum approximation and can be cut off at the lattice constant scale $a$.  As a result, with logarithmic accuracy, the ground-state energy is given by $E_{0}=-gZ^{2}Ry^{*}\ln(1/q_{s}a)$.

For bilayer graphene the range of applicability of the TF result (\ref{screened_Coulomb_real_space}) is further constrained by the condition of validity of the parabolic approximation to the spectrum (\ref{dispersion_law}), so that $n(r)\lesssim n^{*}$ must apply. The outcome is $bZ^{1/3} \ll r \ll bZ$ which is where the potential and density fall off as $1/r^{3}$.  For smooth external potentials created by the gates the TF theory can be applicable in a substantially wider range.  
\begin{figure}
\includegraphics[width=1.0\columnwidth, keepaspectratio]{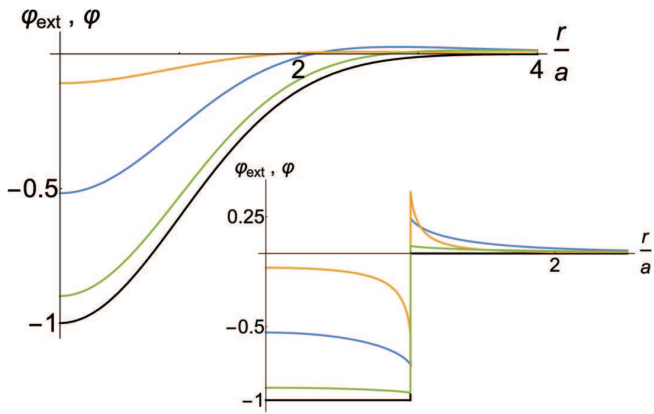} 
\caption{(Color online) External Gaussian potential $\varphi_{ext}\propto-\exp(-r^{2}/2a^{2})$ (bold) and corresponding total potential $\varphi$ (both in arbitrary units) as functions of position for different values of the screening parameter from weak to strong:  $q_{s}a=0.1$ (green), $q_{s}a=1$ (blue), and $q_{s}a=10$ (orange).  The inset:  the same for rectangular well external potential of range $a$. }
\end{figure}  

External potentials that are less long-ranged than the Coulomb potential mimic the effect of neutral impurities.  Since a nearly arbitrary external potential can be designed with the help of the gates, studying the response to a localized potential is also needed.   Here we encounter an effect that is qualitatively different from that due to the Coulomb potential.  Assume there exists a sufficiently localized external potential $\varphi_{ext}(\textbf{r})$ of definite sign.  If $\varphi_{ext}<0$,  it is a potential well that binds holes.  Even though the holes screen the imposed potential, the screening process does not stop here because the holes carry electric charge whose long-range field has to be further screened by creating electrons outside the range of $\varphi_{ext}$ in the amount necessary to guarantee electric neutrality.  We conclude that the bilayer dielectric response to the presence of localized external potential of define sign always involves both electrons and holes; the resulting total potential and density are bound to be of variable sign.  This can be formally seen by assuming that the external potential falls off faster than $1/r$ and taking the $r\rightarrow \infty$ limit in Eq.(\ref{scpotential}).  The outcome is a statement of net zero induced charge ($\int n(\textbf{r})d^{2}r=0$).  As in the Coulomb case, there is no threshold to screening by the electron-hole pairs.  Indeed, let us assume there initially is localized external potential $\varphi_{ext}>0$ and no carriers present.  Since the band gap is zero and there always exists a bound state in two dimensions, the system energy can be lowered by creating an electron-hole pair and binding it to the potential with the electron localized in the neighborhood of the potential and hole being at the periphery of $\varphi_{ext}$;  the hole is needed for neutrality.           

This effect is shown in Fig.1 where we plotted the total potential for the case of a Gaussian potential, $\varphi_{ext}(r) \propto -\exp(-r^{2}/2a^{2})$, for various values of the dimensionless screening parameter $q_{s}a$.  For $q_{s}a\ll1$ the total potential within the well is only slightly smaller (in magnitude) than $\varphi_{ext}$.  As the efficiency of screening increases (i.e. $q_{s}a$ becomes large), the potential goes to zero while the carrier density approaches its classical limit (\ref{classical}).  The inset shows the same effect for the rectangular well potential where additionally one sees "piling up" of the charges at the well boundary which is an illustration of their Coulomb interactions;  the classical density distribution (\ref{classical}) in this case is singular at the well boundary.

This work was supported by US AFOSR Grant No. FA9550-11-1-0297.

\end{document}